\documentstyle[12pt]{article}

\begin{document}

\topmargin 0pt
\oddsidemargin 5mm
\def\bbox{{\,\lower0.9pt\vbox{\hrule \hbox{\vrule height 0.2 cm
\hskip 0.2 cm \vrule height 0.2 cm}\hrule}\,}}
%%%%%%%%%%%%%%%%%%%%%%%%%%%%%%%%%%%%%%%%%%%%%%%%
\def\a{\alpha}
\def\b{\beta}
\def\g{\gamma}
\def\G{\Gamma}
\def\d{\delta}
\def\D{\Delta}
\def\e{\epsilon}
\def\h{\hbar}
\def\ve{\varepsilon}
\def\z{\zeta}
\def\t{\theta}
\def\vt{\vartheta}
\def\r{\rho}
\def\vr{\varrho}
\def\k{\kappa}
\def\l{\lambda}
\def\L{\Lambda}
\def\m{\mu}
\def\n{\nu}
\def\o{\omega}
\def\O{\Omega}
\def\s{\sigma}
\def\vs{\varsigma}
\def\S{\Sigma}
\def\vphi{\varphi}
\def\av#1{\langle#1\rangle}
\def\pa{\partial}
\def\na{\nabla}
\def\hg{\hat g}
\def\un{\underline}
\def\ov{\overline}
\def\cF{{{\cal F}_2}}
\def\Hsl{H \hskip-8pt /}
\def\Fsl{F \hskip-6pt /}
\def\cFsl{\cF \hskip-5pt /}
\def\ksl{k \hskip-6pt /}
\def\pasl{\pa \hskip-6pt /}
\def\tr{{\rm tr}}
\def\tcF{{\tilde{{\cal F}_2}}}
\def\tg{{\tilde g}}
\def\shalf{\frac{1}{2}}
\def\nn{\nonumber \\}
\def\w{\wedge}
%%%%%%%%%%%%%%%%%%%%%%%%%%%

\def\cmp#1{{\it Comm. Math. Phys.} {\bf #1}}
\def\cqg#1{{\it Class. Quantum Grav.} {\bf #1}}
\def\pl#1{{\it Phys. Lett.} {\bf B#1}}
\def\prl#1{{\it Phys. Rev. Lett.} {\bf #1}}
\def\prd#1{{\it Phys. Rev.} {\bf D#1}}
\def\prr#1{{\it Phys. Rev.} {\bf #1}}
\def\prb#1{{\it Phys. Rev.} {\bf B#1}}
\def\np#1{{\it Nucl. Phys.} {\bf B#1}}
\def\ncim#1{{\it Nuovo Cimento} {\bf #1}}
\def\jmp#1{{\it J. Math. Phys.} {\bf #1}}
\def\aam#1{{\it Adv. Appl. Math.} {\bf #1}}
\def\mpl#1{{\it Mod. Phys. Lett.} {\bf A#1}}
\def\ijmp#1{{\it Int. J. Mod. Phys.} {\bf A#1}}
\def\prep#1{{\it Phys. Rep.} {\bf #1C}}

%%%%%%%%%%%%%%%%%%%%%%%%%%%%%

\begin{titlepage}
\setcounter{page}{0}

\begin{flushright}
IASSNS-HEP-97/118 \\
hep-th/9710219 \\
October 1997
\end{flushright}

\vspace{5 mm}
\begin{center}
{\large Matrix Models and String World Sheet Duality }
\vspace{10 mm}

{\large S. P. de Alwis\footnote{e-mail: dealwis@sns.ias.edu, 
dealwis@gopika.colorado.edu}~\footnote{On leave from 
Department of Physics, Box 390,
University of Colorado, Boulder, CO 80309. }}\\
{\em School of Natuaral Sciences, Institute for Advanced Study, 
Princeton NJ 08540}\\
\vspace{5 mm}
\end{center}
\vspace{10 mm}

\centerline{{\bf{Abstract}}}
The scaling limits used recently to derive matrix models, and
a certain analyticity assumption, are invoked to argue that
the agreement between some matrix model calculations and supergravity is
a consequence of string  world sheet duality. 

\end{titlepage}
\newpage
\renewcommand{\thefootnote}{\arabic{footnote}}
\setcounter{footnote}{0}

\setcounter{equation}{0}
In a recent paper  Seiberg\cite{ns} has given a derivation of the 
matrix model\cite{bfss}. However there 
 appears to be some confusion as to whether the argument in \cite{ns}
 effectively bypasses checks on whether gauge theory calculations agree with
 supergravity. The problem stems from the fact that neither in the 
 light-like version nor in the space-like version of the
 matrix model has the connection to
 the supergravity effective action  been directly 
 established as in the case of string theory. In the latter case,
 as is well known, the consistency conditions for the propagation of
 strings results in a background which obeys 
  the equations of supergravity
  with also a systematic prediction as to what the higher
 derivative corrections to Einstein's equations are. There is no
 such demonstration in the case of the matrix model, indeed it has
 not even been put in a covariant form. Hence we believe some
 further clarification, even if only in a certain limited
 area, of the relationship between matrix model calculations
 and supergravity is of some interest.
 To this end we will give an argument using the 
 limit considered in \cite{ns}
 \footnote{This limit has also been
 used by Sen\cite{as} to give a uniform description of matrix models
 and also by J. Maldacena\cite{jm} to discuss the relation to supergravity
 calculations. In fact our discussion will use some of the results of the
 latter paper.}
 an assumption of analyticity, and string 
 theory world sheet duality, to establish that, at least for processes with
 one impact parameter and for 
 which longitudinal (i.e. 11th direction) momentum transfer is zero, 
 the matrix model reproduces
 supergravity.

The action for $N$ Dp-branes  involves in general higher derivative
terms and multiple commutatator terms of the gauge fields ($A$) 
living on the
brane. However in the `gauge theory' limit\cite{ns},\cite{as}\cite{jm}

\begin{equation}\label{limit}
l_s\equiv\sqrt{\a '}\rightarrow 0,~~g\rightarrow 0,
{\rm with}~ g^2_{m}=(2\pi)^{p-2}
gl_s^{p-3},~ A_{\mu},~ {\rm fixed},
\end{equation}
($l_s$ is the string scale, $g$ is the string coupling and $A$ is the 
gauge field ) one gets the matrix model action,
\begin{equation}\label{action}
S =-{1\over 4g^2_{m}}\int_{W_{p+1}}\tr (F_{\a\b}F^{\a\b}+D_{\a}X^iD^{\a}X_i
+[X^i,X^j]^2).
\end{equation}
Note that the indices $\a,\b$ are those tangential to the p-brane and $i,j$
are transverse to it. The  fields were identified in terms of the
ten-dimensional gauge fields in the usual fashion, i.e. 
$F_{\a\b}=[D_{\a},D_{\b}],~D_{\a}=\pa_{\a}-A_{\a},
~ X_i=A_i, F_{\a i}=D_{\a}X_i,
F_{ij}=[X_i,X_j]$. These fields are $U(N)$ matrices and the expression for
the  action includes a $U(N)$ trace.

In the rest of this note we will confine ourselves to zero-branes. 
It is expected that
considerations involving other branes can be obtained in a similar
manner. However it is important to point out that in the limit
(\ref{limit}) only one kind of brane survives. Thus in 10 dimensions
one chooses $p=0$ and only zero-branes will survive with all
other branes being composites of those. This is of course part of
the derivation of matrix models discussed in \cite{ns}, \cite{as}.
Thus our arguments will be valid only in these 
maximally supersymmetric situations.

In matrix model calculations of the forces between branes one
 integrates out
fluctuations around a classical background configuration corresponding to 
the relative positions of the branes\cite{bfss}. Consider $N$ D0
branes with one of them being treated as a probe brane separated 
from the others (which are coincident) by a distance (impact parameter)
$b$ along the 2-axis and moving with a velocity $v$ along the 1-axis.

 Thus we put 
$X^i=B^i+Y^i$, where the background gauge field 
$B^1$ has the element ${vt/ l_s^2}$ on the upper
left-hand corner and zeros everywhere else and $B^2$ has ${b/ l_s^2}$
at the upper left-hand corner and zero everywhere else. The important point
is that the `gauge theory' limit in which $B$ is held fixed as 
$l_s\rightarrow 0$
corresponds to holding $v/ l_s^2$ and ${b/ l_s^2}$ fixed. The term in the 
effective lagrangian coming from the  L-loop
gauge theory diagram with $I$ insertions of the `velocity' background field
 $F\equiv \dot B^1$ then takes the form 
\begin{equation}\label{gauge}
c_{I,L}(N)g^{2L-2}_{m}{F^I\over X^{3L+2(I-2)}}=
c_{I,L}(N){F^2\over g^2_{m}}
\left ({g^2_{m}F^2\over X^{7}}\right )^L
\left ({F\over X^2}\right )^{I-2L-2}.
\end{equation}
This is just the standard loop expansion in the large N limit with the factors 
of $X\equiv B^2$ inserted by dimensional analysis. (See \cite{bbpt}
\cite{jm},
and references therein.)

Let us now consider the string theory calculation of this effective
action to arbitrary order in string
perturbation theory. Firstly in this limit since $g\rightarrow 0$, all
handles (corresponding to string creation and annihilation) are suppressed.
Thus at any order one has an integrand with
 a product of terms corresponding to cylinderical 
world sheets attached to a disc with factors of the form 
 \begin{equation}\label{}
e^{-t(k^2+ M^2)}.
\end{equation}

Here $t$ is some Schwinger parameter and 
$M^2=M_B^2+M_F^2+M_g^2$ is the squared
mass operator, with the different terms on the right hand side being the 
bosonic fermionic and ghost contributions. Explicitly we have \cite{bachas} 

\begin{eqnarray}\label{}
M_B^2&=&\left ( {b\over 2\pi l_s^2}\right )^2+i{\e\over l_s^2}N_0^+
+{i\e (1-\e )-2a_B\over l_s^2}\nn &+&{1\over l_s^2}\sum (non-zero~modes).
\end{eqnarray}

(where $\pi\e=\tanh^{-1}v$ and $N_0^+$ is a zero mode in a light-like
 direction))
 with similar expressions for the fermions and the ghosts. It should
 be stressed that this whole calculation  makes sense only in the
 superstring context\cite{bachas}. The purely bosonic 
 contribution would diverge
 at $v=0$ so that the velocity expansion would make no sense.
 Thus although we will not be making explicit use of supersymmetry
 the fact that we are dealing with the superstring seems
 essential to our considerations. In effect
 the argument indirectly implies the existence of
  a supersymmetric non-renormalization
 theorem. 

In the `gauge theory' limit ($l_s\rightarrow 0$ $b/l_s^2,~\e/l_s^2
\rightarrow v/l_s^2$ fixed), the only
 surviving (non-constant) contribution to the mass operator is 
 \begin{equation}\label{}
 M^2 \rightarrow \left ( {b\over 2\pi l_s^2}\right )^2+{iv\over l_s^2}
 (N_0^+-N^+_0(R))
\end{equation}
where $N_0, N_0(R)$ are certain zero modes coming from the bosonic
and Ramond sectors\cite{bachas}.
The point is that all the massive open string states drop out in the
gauge theory limit and only the BPS states survive\footnote{As this 
paper was being prepared for publication a paper  which also contains
this observation appeared\cite{do}}.

The effective lagrangian of the probe brane as computed from
 open string theory has the general form 
 \begin{equation}\label{string}
L(X,F,g^2_{m},l_s)=\sum_{I,L}c_{I,L}(N,l_sX)
{F^2\over g_{m}^2}\left ({g^2_{m}F^2\over X^{7}}\right )^L
\left ({F\over X^2}\right )^{I-2L-2}.
\end{equation}
In the limit $l_s\rightarrow 0$ keeping $X,F, g_{m}$ fixed this must, by
the previous argument, reduce to the `gauge theory' expression 
(\ref{gauge})
so that $c_{I,L}(N,0)=c_{I,L}(N)$ where the right hand side is the 
coefficient
in the gauge theory. 

Now consider the region $l_sX={b\over l_s}>1$ where the effective action
can be computed in terms of a supergravity action which will contain
an infinite series of higher derivative terms.
The effective action in question is that 
of a brane propagating in the corresponding supergravity background. 
The action
of such an object may be written as 
 \begin{equation}\label{}
S=-{1\over gl_s}\int dt e^{-\phi}\sqrt{\det g}+{1\over l_s}\int C
\end{equation}
where the zero mode of the dilaton has been explicitly factored out and
 the last
term is the coupling of the R-R  field. From the closed
string point of view as long as the background curvature is small
($l_s^2R<<1$) which is the case for large $l_sX>1$, there is a meaningful
expansion for $g,\phi.C$ that is consistent with closed string
propagation in powers of $l_s$. These fields will be solutions of the
all orders in $l_s$ expansion in supegravity Lagrangians.
 In terms of the original parameters
$F$ and $X$ however this version of the effective action must 
have the same expasion as (\ref{string}) except that the coefficients
$c_{I,L}(N,l_sX)$  are now replaced by $c_{I,L}^{SG}(N,l_sX)$ the 
superscript $SG$ denoting the fact that they are to be computed from
the (infinite series of) supergravity actions. Now in the region
$l_sX>1$  one has from the usual closed string argument, $c_{I,L}(N,l_sX)
=c_{I,L}^{SG}(N,l_sX)$. Now we have to make a crucial assumption that
these functions are analytic. Clearly since the supergravity expansion
must break down at $l_sX=1$ there must be a pole at this value
on the real axis. However there is no reason to expect any cuts etc which
would prevent the analytic continuation of this equation to the region inside
the circle $|l_sX| =1$ and in particular to the point $l_sX=0$. Thus
assuming analticity we have the result,
\begin{equation}\label{equal}
c_{I,L}(N)=c_{I,L}(N,0)=c_{I,L}^{SG}(N,0)
\end{equation}
where the first equality  was established earlier.

For the diagonal terms ($I=2L+2$) the coefficient $c_{I,L}^{SG}(N,0)$
can be calculated from the lowest order in $l_s$ supergravity action.
This is because on dimensional grounds the higher derivative terms 
will not contribute to these terms \cite{bbpt}.
Thus  if one puts in the explicit 
classical
supergravity solution \cite{hs},\cite{jm}, we get
 \begin{equation}\label{}
S=-{1\over (2\pi )^2g_m^2l_s^4}
\int dt f^{-1}(\sqrt{1-fv^2}-1)
\end{equation}
where 
\begin{equation}\label{}
f=1+{k\over l_s^4},~~~k\equiv {cg_m^2N\over X^7}
\end{equation}
with c a known constant.
Now taking the gauge theory limit one gets (see for
example \cite{jm} and references therein) 
\begin{equation}\label{limaction}
L(X,F,g^2_{m},l_s=0)=-{1\over g^2_{m}}k^{-1}(\sqrt{1-kF^2}-1).
\end{equation}

But by our previous argument and in particular equation
(\ref{equal}) this is the same object that 
was calculated  in the gauge theory. Hence the two 
expressions must be the same. 
This appears to be a generalization of the old result that
the closed string (and hence classical gravity) appears  as a 
quantum effect in open string perturbation theory.
As has been pointed out in \cite{bfss} this object has an
 M-theory interpretation. In particular as shown in \cite{bbpt}  
the metric dialton and RR field in (\ref{limaction}) can be
regarded as being obtained from the null reduction of 
a M-theory metric due to a supergraviton source.

It is useful at this point to discuss the units in which various 
physical quantities in the theory are being defined. While a choice
of units is obviously not going to change the physics,
a convenient choice will clarify the aspects of the physics that 
we wish to study better than some other choice. Since what seems to
emerge from the study of matrix models is  11-D supergravity
it is natural to set the (classical) eleven dimensional Newton constant
equal to one. (This is particularly  useful if one wishes to 
study quantum effects around classical solutions).
 With the velocity of light being set equal to
one also, we keep
Planck's constant $\h =l_P^9$. i.e. the parameter that defines the 
semi-classical (quantum loop) expansion is 
 the ninth power of the Planck length. This is the natural system of
 units to use in any discussion of quantum (semi-classical) 
 corrections (such as Hawking radiation) to  solutions
 of classical gravitational field equations (such as black holes).
 We then have the following formulae.
 \begin{eqnarray}\label{lim}
16\pi G_N=1,~c=1, & & \h\equiv l_P^9,~~l_{m}^{-3}\equiv (2\pi g_{m})^2\nn
g^2={l_P^3\over l_{m}^3}={\h^{1\over 3}\over l^3_{m}},& &l_s=
l_P^{1\over 2}l_{m}^{1\over 2}=\h^{1\over 18}l_{m}^{1\over 2} \nn
{b\over l_P}=Xl_{m}={x_m\over l_m},& & {v\over l_P}=Fl_{m}
\end{eqnarray}
 The above formulae clarify the relation between the matrix model
 quantum mechanics (characterized by the Yang-Mills length scale
 $l_{m}$) and the 11D or target space quantum mechanics characterized
 by $l_P\equiv \h^{1\over 9}$. The limit considered in \cite{ns},\cite{as},
 \cite{jm}, corresponds to taking the dimensionless ratio $l_P\over l_{m}$
 to zero. As one sees from (\ref{gauge}) the condition for
 the validity of the loop expansion in Yang-Mills quantum mechanics is that
 the dimensionless number $l_{m}X={b\over l_P}>>1$. In other words measured
 in 11D  Planck  units the impact parameter must be large, although measured
 in string units it is going to zero.

The rescaling to gauge theory variables (and the above choice of units)
 also clarifies the issue  of
whether higher derivative terms  in the 11-D supergravity
action (which should be there on general grounds as quantum corrections
- see for example \cite{mg}) are reproduced by matrix model calculations.
At first sight the answer seems to be negative. Let us write 
the semi-classical expansion (which in our units is the same
 the low-energy expansion) of the action as 
\begin{eqnarray}\label{}
I&=& {1\over l_P^9}\int\sqrt g R +...\nn
&+&{1\over l_P^7}\int d^{11}\sqrt g ``R^2"+.... 
\end{eqnarray}  
In the above the first line is the classical 11-D action while the second
line denotes all possible $R^2$ terms etc coming from quantum effects.
The  limit $l_P\rightarrow 0$ in this action would 
just appear to pick up the leading classical term which would seem to 
imply that the matrix model just gave only classical supergravity.
However we also need to rescale the coordinates (corresponding to
the last line of (\ref{lim}) such that 
\begin{equation}\label{}
x\rightarrow x_{m}=l_{m}^2X={l_{m}\over l_P}x .
\end{equation}
Since the limit is taken with $l_{m}$ fixed this means that small distances
in the original variables become large distances in the new variable.
 Thus the action is now 
\begin{eqnarray}\label{}
I&=& {1\over l_{m}^9}\int d^{11}x_{m}\sqrt g R_{m} +...\nn
&+&{1\over l_{m}^7}\int d^{11}x_{m}\sqrt g ``R^2"_{m}+.... 
\end{eqnarray}  
Thus the quantum expansion parameter is now $l_{m}$ which is fixed and
the matrix model is expected to pick up all the higher order terms.
Thus any non-vanishing non-diaganol terms in the matrix model calculation
should correspond to the higher-derivative terms in the supergravity
effective action.
The two theories one with Planck's constant $l_P^9$ the other with 
Planck's constant $l_m^9$ appear to be
  equivalent to the two theories, the auxiliary one with 
  Planck mass $\tilde M_P$ 
 and the other with Planck mass $M$,
introduced by Seiberg\cite{ns}.

We should stress however that the actual argument made above
applies only to zero momentum transfer processes in the 11 direction
(since it depends on string theory arguments).
We believe that this is consistent with the argument in \cite{ns} that
the finite N light-like compactified M-theory is equivalent to the 
limit (\ref{limit}) of string theory.  It seems to us therefore
that further calculations such as that of Polchinski and Pouliot\cite{pp} 
provide non-trivial checks of the validity of the matrix model.

Before we conclude we should stress that the arguments of this
paper can be interpreted as just a statement about ten-dimensional
string theory (as in \cite{dkps})
 without any reference whatsoever to \cite{bfss}
and the subsequent developments. Thus one can think
about the result as being just a statement relating the calculation
 of a D0-brane effective action from two different
 perpectives. Obviously however they are
of current interest because of the possible connection to
M-theory that was first pointed out in \cite{bfss}.

Finally we should comment on two papers\cite{dr},\cite{do} that appeared 
as this note
was being prepared for publication. The first  involves the scattering
of three gravitons to three gravitons and so it depends on
 two impact parameters.
The arguments above apply explicitly  only to the  one impact parameter 
situation. In particular on the closed string side of the argument
we have used analyticity in impact parameter space (which is equivalent
to analycity in the complex angular momentum place\footnote{I'm grateful to
V.P. Nair for pointing this out to me.}). In the two impact parameter
case  the singularity structure is clearly more complicated it is 
possible that  this analytic
continuation is invalid.
Similar remarks would apply to the relation 
of this work to
\cite{do} where the matrix theory is tested on an ALE space. 

{\bf Acknowledgments:} I would like to thank Dan Kabat and 
Nati Seiberg for useful discussions and comments on the manuscript.
I'm also grateful to, Finn Larsen, Joe Polchinski, and Ashoke Sen
for comments which led me to clarify the assumptions behind my argument
and to V.P. Nair and Peter Orland and Tom Banks
for comments on the relevance of  Regge
poles and cuts.
  I would also like to thank
Edward Witten for hospitality at the Institute for
Advanced Study and  the Council on  Research and Creative Work
 of the University of Colorado
for the award of a Faculty Fellowship. This work is partially supported by
the Department of Energy contract No. DE-FG02-91-ER-40672.

%%%%%%%%%%%%%%%%%%%%%%%%%%%%%%%%%%%%%%%%%%

%%%%%%%%%%%%%%%%%%%%%%%%%%%%%%%%%%%%%%%%%%%%%%%%
\end{document}